\documentclass[aps,reprint,prb,superscriptaddress,amsmath,amssymb,floatfix]{revtex4-2}

\usepackage[utf8]{inputenc}
\usepackage{graphicx}
\usepackage[colorlinks,allcolors=blue]{hyperref}

\newcommand{\cre}[1]{{b}^\dagger_{#1}}
\newcommand{\ann}[1]{{b}_{#1}}
\newcommand{\num}[1]{{n}_{#1}}

\begin{document}

\title{
Scale-invariant phase transition of disordered bosons in one dimension}

\author{Tanul Gupta}
\affiliation{University of Strasbourg and CNRS, CESQ and ISIS (UMR 7006), aQCess, 67000 Strasbourg, France}

\author{Guido~Masella}
\affiliation{QPerfect, 23 Rue du Loess, 67000 Strasbourg, France}

\author{Francesco~Mattiotti}
\affiliation{University of Strasbourg and CNRS, CESQ and ISIS (UMR 7006), aQCess, 67000 Strasbourg, France}

\author{Nikolay V. Prokof'ev}
\affiliation{Department of Physics, University of Massachusetts, Amherst, MA 01003, USA}

\author{Guido~Pupillo}
\affiliation{University of Strasbourg and CNRS, CESQ and ISIS (UMR 7006), aQCess, 67000 Strasbourg, France}
\affiliation{QPerfect, 23 Rue du Loess, 67000 Strasbourg, France}

\begin{abstract}
The disorder-induced quantum phase transition between superfluid and non-superfluid states of bosonic particles in one dimension is generally expected to be of the Berezinskii-Kosterlitz-Thouless (BKT) type. Here, we show that hard-core lattice bosons with power-law hopping decaying with distance as $1/r^\alpha$ with finite integral over space -- corresponding in spin language to a $XY$ model with power-law couplings -- undergo a non-BKT continuous phase transition instead. We use exact quantum Monte-Carlo methods to determine the phase diagram for different values of the exponent $\alpha$, focusing on the  regime $\alpha > 2$. We find that the scaling of the superfluid stiffness with the system size is scale-invariant at the transition point for any $\alpha\leq 3$ -- a behavior incompatible with the BKT scenario and typical of continuous phase transitions in higher dimension.
By scaling analysis near the transition point, we find that our data are consistent with a correlation length exponent satisfying the Harris bound $\nu \geq 2$ and demonstrate a new universal behavior of disordered bosons in one dimension.
For $\alpha>3$ our data are consistent with a BKT scenario where the liquid is pinned by infinitesimal disorder.
\end{abstract}

\maketitle
Bosonic particles with local interactions in one dimension (1D) are described by a universal harmonic theory, known as Luttinger liquid (LL). The latter corresponds to quantized superfluid hydrodynamics (including instantons) and is fully characterized by the superfluid velocity, $v=\sqrt{Y_s/\kappa}$, and LL parameter, $K=\pi \sqrt{\kappa Y_s}$, with $\kappa$ the compressibility and $Y_s$ the superfluid stiffness. Diagonal disorder induces an instability in LL 
towards a non-superfluid Bose glass (BG) phase -- 
a compressible insulator displaying exponential decay of off-diagonal correlations. 
In their seminal paper \cite{Giamarchi1988}, Giamarchi and Schulz found by means of a perturbative renormalization group (RG) analysis that the LL-BG transition is of the Berezinskii-Kosterlitz-Thouless (BKT) type that takes place at the universal value
$K=K_{c} = 3/2$ (this result holds at the two-loop level  \cite{Ristivojevic2014}). In the
strong-disorder limit, real-space RG treatments  \cite{Altman2008, Altman2010} and the ``scratched-XY'' criticality \cite{PolletNikolay2014} also predict a BKT-type transition 
but at a non-universal value of $K_c>3/2$. These considerations 
exhaust known scenarios for the disorder-induced superfluid to non-superfluid phase transitions in 1D. 

In this work, we consider the disorder-induced localization transition in 1D superfluids of bosons with power-law hopping decaying with distance as $1/r^\alpha$. We utilize numerically exact large scale Quantum Monte-Carlo simulations based on the Worm Algorithm \cite{prokof1998exact} to determine the ground-state superfluid phases and phase transitions for different values of  $\alpha>2$. We find that the superfluid phases can be approximately characterized by an effective LL parameter $K$ that reproduces the decay of correlation functions. However, contrary to existing theories, we find that the disorder-induced quantum phase transition is generically scale-invariant and incompatible with the BKT scenario with the effective $K_c\leq 3/2$ for all $\alpha\leq 3$. As far as critical exponents are concerned, the data is consistent with the correlation length exponent satisfying the Harris bound $\nu \geq 2$ for all values of $\alpha\leq 3$. Thus, our results reveal a new universal behavior of bosons with power-law hopping in one dimension with finite integral over space. For $\alpha>3$ our results are instead consistent with a scenario where the superfluid is pinned by an infinitesimal disorder in the thermodynamic limit, similar to a BKT-like scenario for hard-core particles with short-range coupling. Our predictions are directly relevant for experiments with dipolar atoms and molecules, exciton materials, and cold ions.  

We consider the following 1D lattice Hamiltonian for hard-core bosons 
\begin{equation}\label{eq:hamiltonian}
    \mathcal{H} =
    -t \sum_{i < j} \frac{a^\alpha}{|r_{ij}|^\alpha}
    \left[ \cre{i}\ann{j} + \text{H.c.} \right]
    + \sum_i \epsilon_i \num{i} , \qquad (n_i \le 1).
\end{equation}
We employ standard notations for bosonic creation and annihilation operators on site $i$ and restrict the maximal occupation number, $\num{i} = \cre{i}\ann{i}$, to unity. The nearest-neighbor hopping amplitude, $t$, and the lattice spacing, $a$, are taken as units of energy and length, respectively.
We choose random on-site energies $\epsilon_i$ uniformly distributed between $-W$ and $W$, and check that a different (gaussian) choice of distribution does not affect the results.
In spin language, Eq.~\eqref{eq:hamiltonian} is equivalent to an $XY$ Hamiltonian with power-law exchange couplings, which, in the absence of disorder, can be realized in experiments
with cold polar molecules \cite{Yan2013}, trapped ions
\cite{Richerme2014,Jurcevic2014, feng2022continuous} and Rydberg atoms
\cite{Zeiher2017,Barredo2015,Orioli2018, browaeys_lahaye_2020, Semeghini2021, chen2023continuous} (the latter can also be disordered \cite{Leseleuc2019}). 

For ideal system with $W/t=0$, the spectra and low-energy phases of Hamiltonian~\eqref{eq:hamiltonian} have been investigated by a variety of approaches. Using linear spin-wave theory, Ref.~\cite{Roscilde2017} identified $\alpha>3$ as a regime where main properties reproduce those observed in the 
$\alpha= \infty$ limit of finite-range interactions;
$1 < \alpha < 3$ as an intermediate regime with 
the $XY$ phase characterized 
by a continuously varying dynamical exponent $z = (\alpha -1)/2$
(it governs the $k \rightarrow 0$ limit of the dispersion relation);
and $\alpha < 1$ as a long-range regime with 
dispersionless excitations and properties similar to the infinite-range $\alpha = 0$ case in the thermodynamic limit. 
In this harmonic approach, $\alpha=3$ is the boundary between
the intermediate and short-range regimes. 
Using a bosonization approach supplemented by an RG analysis, Ref.~\cite{maghrebi_gong_gorshkov_2017} predicts that power-law couplings are relevant in the RG sense for $\alpha < 3 - 1/(2K)$, with $K>1$ to be determined numerically for each given $\alpha$. In the following, we study the ground-state superfluid phases and phase transitions of Eq.~\eqref{eq:hamiltonian} for $\alpha>2$ using large scale path-integral quantum Monte-Carlo
simulations based on the Worm algorithm \cite{prokof1998exact}. Without loss of
generality, we focus on the particle density $\rho=1/2$.

\begin{figure}[t]
    \includegraphics[width=\columnwidth]{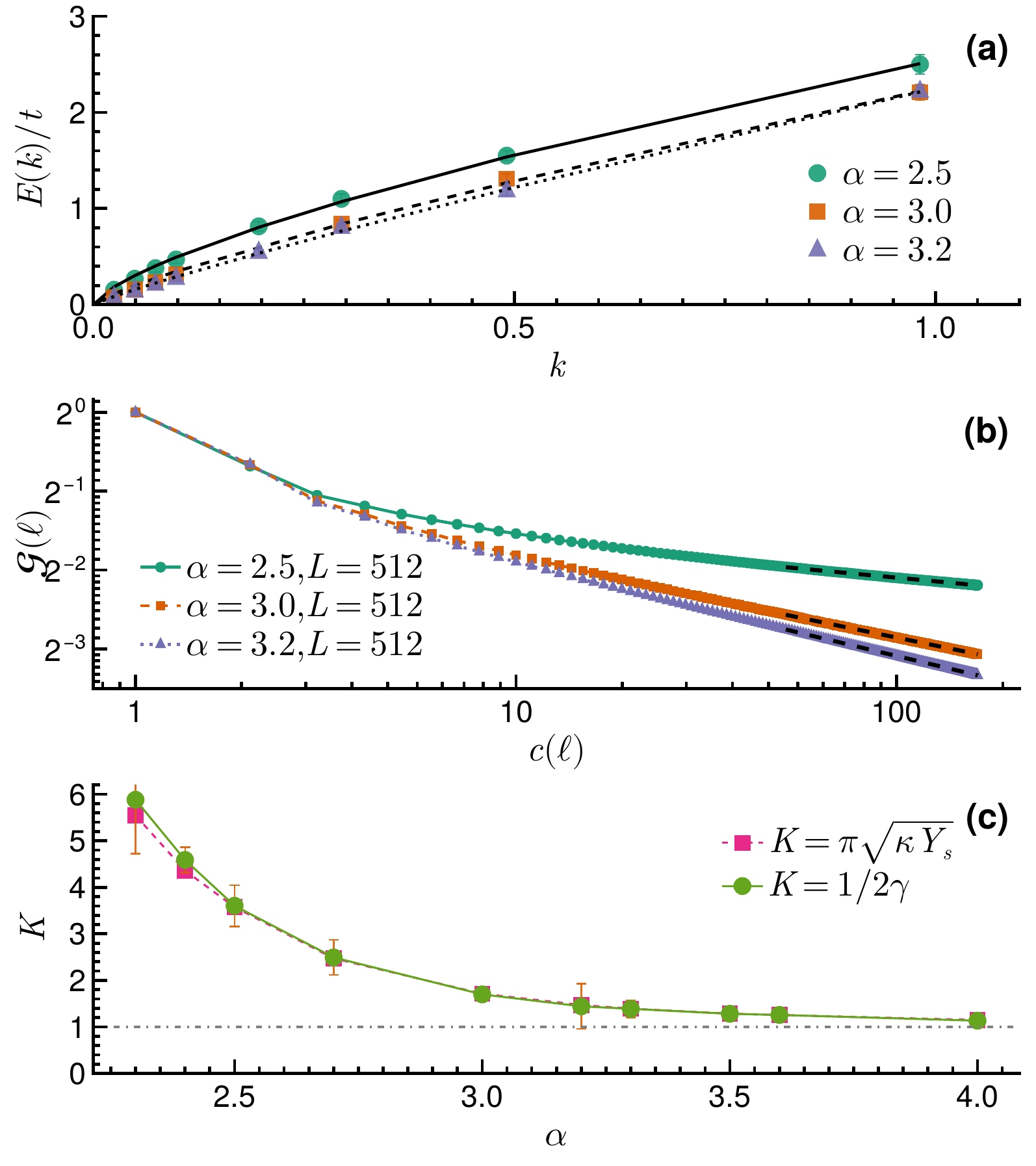}
    \caption{Characterization of the superfluid phase for $W/t=0$: (a) Dispersion relations $E(k)$ vs $k$ for $\alpha=2.5$, 3.0 and 3.2 chosen in the intermediate and short-range regimes, respectively (see text) for $L=256$. (b) Single particle density matrix  $\mathcal{G}(\ell) $ vs chord distance $c(\ell)=\sin(\pi \ell/L)/\sin(\pi/L)$ showing an algebraic decay for all $\alpha$ for $L=512$. Dashed lines indicate the best fit with $A\cdot c(\ell)^{-\gamma}$ where $A,\gamma$ are fitting parameters. (c) Numerical evaluation of the Luttinger liquid parameter $K$ as a function of $\alpha$ from the power-law decay $\mathcal{G} \propto \ell^{-1/(2K)}$ (green dots) and from the relation $K=\pi \sqrt{\kappa Y_s}$ (red squares) extrapolated to the thermodynamic limit via a polynomial scaling in $1/L$.}
 \label{fig:1}
\end{figure}

We start our analysis by first characterizing the bosonic liquid in the absence of disorder ($W/t=0$). Figure \ref{fig:1}(a) shows the dispersion relation $E(k)$ vs $k$ 
for three values of $\alpha=2.5,\, 3.0,$ and $3.2$ where $k$ is the quasi-momentum. It was deduced numerically from spectral peaks after analytic continuation of the imaginary-frequency dynamic structure factor \cite{prokofev_ac_2013, prokofev_ac_2017}. The chosen values of $\alpha$ correspond to values in the expected intermediate ($\alpha=2.5$), boundary ($\alpha=3.0$) and short-range ($\alpha=3.2$) limits of the spin-wave analysis, respectively. The dispersion relation is non-linear in $k$ for $\alpha =2.5$ (dots) and the data can be fit well 
by $E(k)\sim k^{z_*}$, with $z_* \simeq 0.74$, in good agreement with the $z = 0.75$ prediction of spin-wave analysis (continuous black line). In the short-range regime, instead, the dispersion relation is consistent with the linear law and a small negative quadratic contribution, also in agreement with literature.
We checked that $E(k)$ for different system sizes agree with each other for the same values of $k$. [The dispersion relation in the superfluid phase for $\alpha = 2.5$ in the presence of disorder is sub-linear, as presented in the Supplemental Material~\cite{Note1}.]

For a 1D superfluid ground state, the single-particle density matrix 
$\mathcal{G}(\ell)=\langle \, \cre{i}\ann{i+\ell }\rangle$ 
is expected to show an algebraic decay with the  distance $\ell$ with diverging integral over space. Our data for $\mathcal{G}$ are shown in Fig. \ref{fig:1}(b), for the same values of $\alpha$ as in panel (a) 
for a system with $L=512$ sites and inverse temperature $\beta=L/t$.
We observe algebraic decay $\mathcal{G} \sim \ell^{-\gamma}$ for all $\alpha$. 

\begin{figure*}
\includegraphics[width=\linewidth]{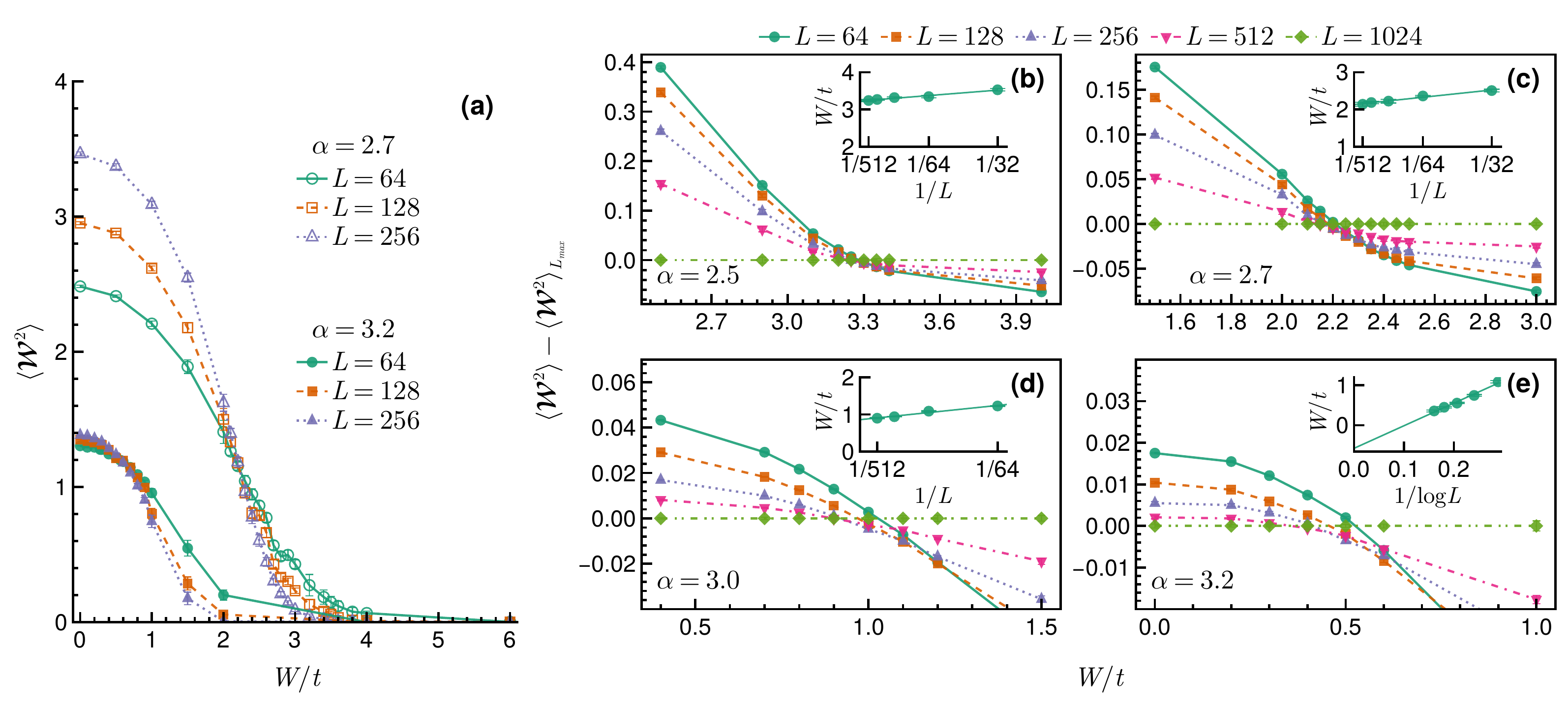}
    \caption{Characterization of the superfluid to non-superfluid phase transition: (a) Mean-square winding number $\langle \mathcal{W}^2 \rangle$ vs disorder strength $W/t$ for $\alpha=2.7$ (empty symbols) and $3.2$ (full symbols) for system sizes $L=64$, $128$, $256$. (b)-(e) Zoom-in on the area near phase transitions for $\alpha=2.5$, $2.7$, $3.0$ and $3.2$, showing crossing between the curves; the curve corresponding to the largest size $L=1024$ is subtracted from all data for clarity. Vertical error bars indicate the estimated uncertainty from the Monte Carlo simulations and disorder-averages. Insets: Finite-size scaling of crossings points between curves for system sizes $L_1$ and $L_2=2L_1$ as a function of $L=L_1$.
    }
    \label{fig:2}
\end{figure*}
Despite the non-linear dispersion relation demonstrated above, we attempt a comparison with expectations from LL theory by extracting 
an effective LL parameter $K$ as a function of $\alpha$ from two standard methods:  the power-law decay $\mathcal{G} \sim \ell^{-\gamma}$  via the bosonization relation $\gamma=1/(2K)$ (green dots) and the relation $K=\pi \sqrt{\kappa Y_s}$ (red squares). Both 
$\kappa$ and $Y_s$ can be conveniently computed by quantum Monte Carlo through mean-square particle, $N$, and winding number, $\mathcal{W}$, fluctuations using the Pollock–Ceperley 
relation $Y_s = L\langle \mathcal{W}^2 \rangle/\beta$. Figure \ref{fig:1}(c) shows that the two methods produce similar estimates of $K$ for all $\alpha$, within the 
error bars, which is surprising, given the non-linear dispersion relation demonstrated above. Moreover, $K$ decreases monotonically and 
continuously with $\alpha$ from a large value $K\gtrsim 5$ at $\alpha \sim 2.3$ to $K\approx 1$ at $\alpha=4$. Within an approximate LL scenario, this behavior may be explained by the fact that power-law hopping in Hamiltonian \eqref{eq:hamiltonian} allows for large-scale particle exchanges for small enough $\alpha < 3$, mimicking the behavior of soft-core bosons, for which one can easily get $K\gg 1$. 
The $K=1$ value (dashed dotted line) corresponds to the 
short-range case of hard-core bosons with the nearest neighbor hopping, a limit that is here asymptotically approached at $\alpha> 3$ \cite{GiamarchiBook}.
In the following we analyse the situation at finite disorder 
strength and, in particular, explore the nature of the transition point, which is expected to be of the BKT type for Luttinger liquids. 
However, we note that this may not be the case here: BKT transition and its asymptotically exact RG flow are rooted in logarithmic interactions between vortex excitations. The latter originates from the kinetic energy of the flow around vortexes $E \sim \int (n_s/m)  dr/r$, where $m$ is the particle mass and $n_s$ the superfluid density. The single particle spectrum in our model is not parabolic and formally corresponds to a scale dependent ``mass'' $m(r)\sim r^{3-\alpha}$, implying that vortexes in the superfluid phase should be bound by a power-law, not logarithmic, potential. It may thus be expected that BKT physics no longer applies for $\alpha\lesssim 3$.

\begin{figure}[b]
\includegraphics[width=\columnwidth]{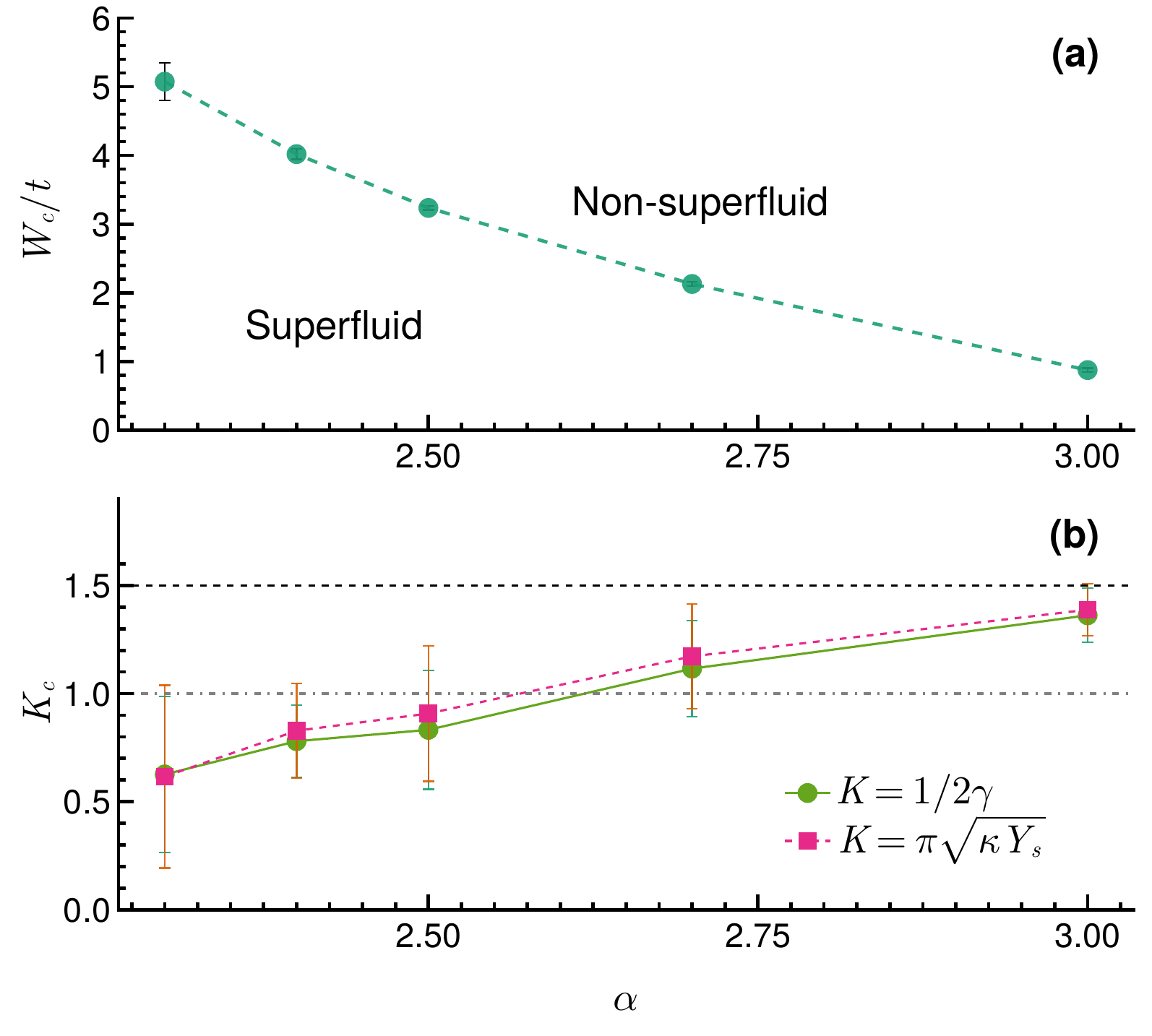}
    \caption{(a) Phase diagram, $W_c$ vs $\alpha$, of the superfluid and non-superfluid quantum phases
    for model (\ref{eq:hamiltonian}). 
    (b) Critical values of LL parameter $K_c$ at $W_c$ vs $\alpha$, 
    as estimated from the power-law decay of $\mathcal{G}$ (green dots) and from $K=\pi \sqrt{\kappa Y_s}$ (red squares) for $L=256$.}
    \label{fig:3}
\end{figure}

We characterize the transition via the winding number fluctuations $\langle \mathcal{W}^2 \rangle$ since they are a scale invariant quantity, differently from the superfluid stiffness $Y_s$.
Figure \ref{fig:2}(a) shows the evolution of superfluid properties measured by $\langle \mathcal{W}^2 \rangle$ with disorder, $W/t$,
for two example cases $\alpha=2.7$ (empty symbols) and 3.2 (full symbols) and for several values of $L=64,128,256$. 
In both cases, $\langle \mathcal{W}^2 \rangle$ decrease monotonically with increasing $W/t$, until they reach near zero
values. This behavior signals the transition between the superfluid and non-superfluid states. In the short-range case $\alpha=3.2$,
the behavior at larger values of disorder is reminiscent of what is expected for a BKT transition when 
in the infinite system  $\langle \mathcal{W}^2 \rangle$ displays a jump to zero at the critical point \cite{GiamarchiBook}. However, surprisingly, for $\alpha=2.7$ there is a clear crossing point of $\langle \mathcal{W}^2 \rangle$ around $W/t\sim 2$. This is inconsistent with the BKT criticality and is, instead, 
a signature of continuous scale-invariant phase transitions. 
This fact can be used to pinpoint the critical disorder strength $W_c$ where superfluidity is lost by the crossing point of 
$\langle \mathcal{W}^2\rangle$ -vs- $W$ curves for different values of $L$. The panels (b)-(e) in Fig.~\ref{fig:2} present data in the vicinity of transition points for $\alpha=2.5, 2.7, 3.0$ and $3.2$ using $\beta=L/(8t)$ [even for $\alpha=3.2$ our temperature is a factor of two smaller 
than the lowest phonon mode].
Crossing points are very pronounced in (b) and (c) for intermediate exponents $\alpha$, leaving no  doubt that we are dealing with generic continuous transitions at $W/t=3.24(5)$ for $\alpha=2.5$ and at $W/t=2.12(5)$ for $\alpha=2.7$. 
The crossings appear to persist when transitioning to the short-range regime $\alpha\gtrsim3$, see panels (d) and (e) in Fig.~\ref{fig:2} with crossings around $W/t=0.87(5)$ for $\alpha=3.0$
and around $W/t=0.5(5)$ for $\alpha=3.2$, contrary to all expectations.
However a careful finite-size scaling up to large system sizes $L=1024$ shows that the transition point for $\alpha>3$ scales to $W/t \rightarrow 0$ in the thermodynamic limit, see Inset in Fig. \ref{fig:2}(e), 
implying the absence of a continuous phase transition at finite $W$ in the thermodynamic limit.
This is different from $\alpha \leq 3$, where the transition point scales to a finite value of $W/t$, see Insets in Fig.~\ref{fig:2}(b-d). 
The breakdown of the BKT scenario for all values $2<\alpha\leq 3$ in Eq. \eqref{eq:hamiltonian} is surprising and is the main result of this work. 

Figure \ref{fig:3}(a) summarizes the ground state phase diagram of Hamiltonian \eqref{eq:hamiltonian} in terms of $W_{\rm c}$ and $\alpha$. Here, for each $\alpha\leq 3$, the critical point $W_{\rm c}$ is determined from the scale-invariant crossing point as described above. The critical disorder strength  $W_{\rm c}/t$ decreases monotonically from a large value $W_{\rm c}/t \sim 5.1$ to $\sim0.9$ for $\alpha = 3$. For $\alpha > 3$ the transition in the thermodynamic limit occurs at  $W_{\rm c}/t = 0^+$. 
The limiting value $W_{\rm c}/t = 0^+$ would correspond to the strictly short-range limit of hard-core bosons with short-range hopping, which are known to be localized by an infinitesimal disorder \cite{GiamarchiBook}.

Figure \ref{fig:3}(b) shows the critical LL parameter $K_c$ computed at $W_{\rm c}/t$ for each value of $\alpha$ assuming that the approximate LL scenario properly describes the system. We find that, for $\alpha \lesssim 3$, $K_c$ remains smaller than the critical BKT value of $3/2$ for short-range hopping models with weak disorder \cite{GiamarchiBook}, confirming that LL theory should not be used to describe the localization transition in the whole range $\alpha \leq 3$. In contrast, for $\alpha>3$  our results are in agreement with conclusion that ideal systems with $K<3/2$ are ultimately pinned by disorder, leading to an insulating BG phase for any finite value of $W/t$.

\begin{figure}
\includegraphics[width=\columnwidth]{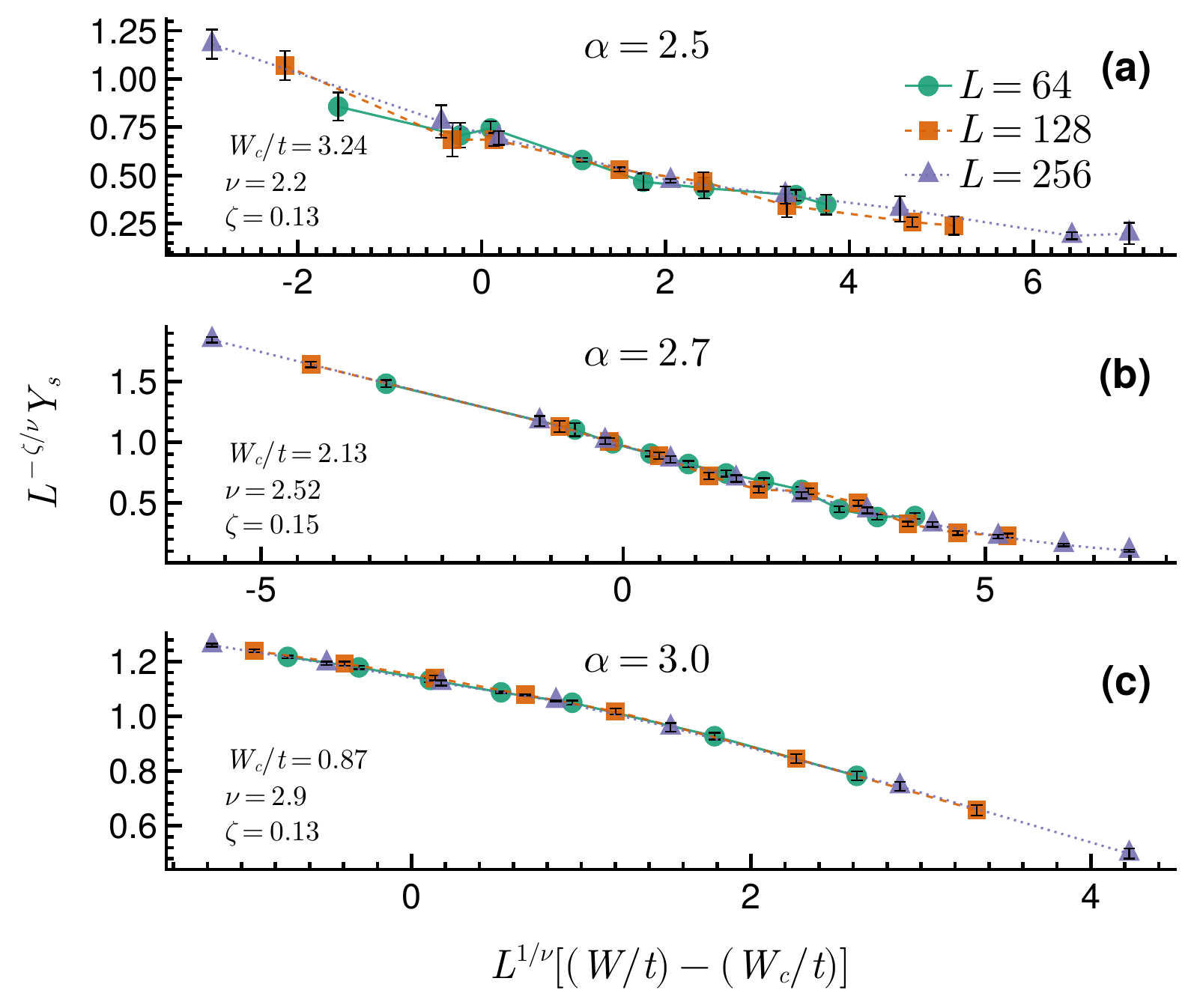}
    \caption{(a)-(c) Data collapse for the scaled superfluid stiffness $L^{-\zeta/\nu}Y_s$ vs $L^{1/\nu}[(W /t) - (W_c/t)]$ for $\alpha=2.5$, $2.7$ and $3.0$ using $L=64$, $128$, $256$. 
    The fitted values of the correlation length exponent $\nu$ and 
    $\zeta$ are reported directly in the figure. They satisfy 
    $ \nu \gtrsim 2$  for all $\alpha$.}
    \label{fig:setup}
\end{figure}

\begin{figure} \includegraphics[width=\columnwidth]{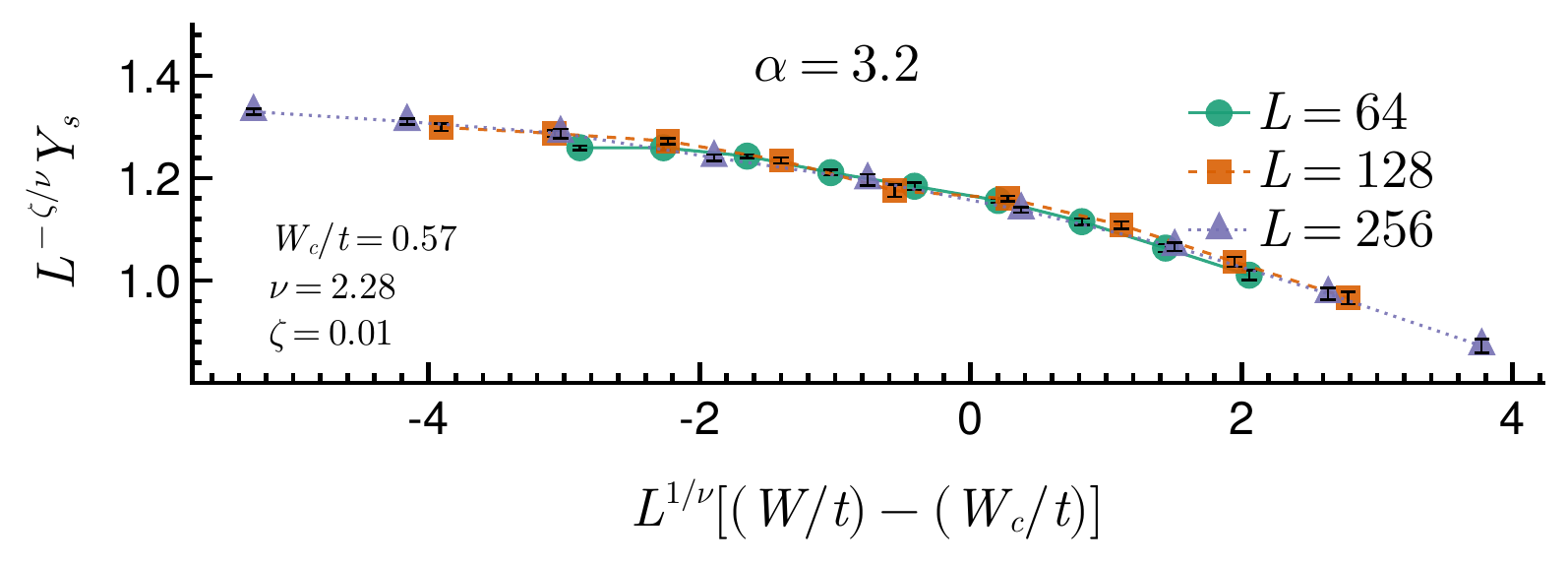} \caption{Data collapse for the scaled superfluid stiffness $L^{-\zeta/\nu}Y_s$ vs $L^{1/\nu}[(W /t) - (W_c/t)]$ for $\alpha=3.2$ using $L=64$, $128$, $256$.     The fitted values of the correlation length exponent $\nu$ and     $\zeta$ are reported directly in the figure. Unlike for $\alpha<3$, this scaling corresponds here to a finite-size effect, as the transition is not located at $W_c/t =0.57$ and slowly shifts to $W_c/t=0$, see Fig. \ref{fig:2}(e).    } \label{fig:setup1} \end{figure}
We complete our characterization of 
quantum phase transition in Fig. \ref{fig:3}(a)
by determining the correlation length exponent $\nu$ 
using data collapse analysis near the critical points \cite{newmanb99}. For each $\alpha$, the results of Monte Carlo simulations are rescaled by $L^{-\zeta/\nu}Y_s$
and collapsed on a single master curve using $L^{1/\nu}[(W /t) - (W_c/t)]$ as a variable. Critical values $W_c/t$ are taken from the crossing points in Fig. \ref{fig:2}, while $\nu$ and $\zeta$ are treated as 
fitting parameters and obtained using a Nelder-Mead algorithm \cite{Nelder1965} with a cost function based on the Kawashima-Ito-Houdayer-Hartmann quality metric \cite{Kawashima1993, Houdayer2004}.
Example results for $\alpha=2.5$, $2.7$, $3.0$ and $\alpha=3.2$ are shown in Figs. \ref{fig:setup} and \ref{fig:setup1}, respectively. We observe good collapse of all data near the critical points 
for $\alpha\ge 2$, 
and the obtained correlation length exponents always satisfy the so-called Harris bound $\nu \gtrsim 2$, see \cite{harris1974effect}. 
In fact, this result is expected from general arguments for a large class of $d$-dimensional disordered systems where an appropriately defined correlation length diverges 
\cite{Chayes1986}. 
Data collapse for $\alpha>3$ using $W_c/t=0.57$ is 
a finite-size effect given that this value of $\alpha$ is close to the boundary between the intermediate and short-range regimes and crossing points slowly shift to zero with increasing the system size.
However, this effect will likely be observed in experiments dealing with finite systems.

In conclusion, we have demonstrated that the disorder-induced superfluid to non-superfluid quantum phase transition for models with power-law hopping 
is a scale-invariant transition if $2<\alpha \leq 3$, ruling out the expected BKT scenario for interacting one-dimensional bosons in this regime. Our work opens up multiple other research directions, including whether the finite-temperature BKT scenario is generally inconsistent with power-law hopping models also in two dimensions \cite{sbierski2023magnetism, diessel_2023, DefenuLongRange2023}. Another open question is the nature of the non-superfluid quantum phase for general values of $\alpha$. In Ref. \cite{Masella2022} it was conjectured that for $\alpha=3$ this phase is
a non-superfluid Bose metal phase with finite zero-frequency optical conductivity and algebraic decay of correlations. It is an open question whether similar behavior can be found for other $\alpha$ values. Our predictions should be directly testable in experiments for $XY$ models realized via internal excitations of cold dipolar atoms and molecules, cold ions chains, and Rydberg atoms. 

\begin{acknowledgments}
We gratefully acknowledge discussions with Jerome Dubail. This research has received funding from the European Union’s Horizon 2020 research
and innovation programme under the Marie Sk{\l}odowska-Curie project 955479 (MOQS), the Horizon Europe programme HORIZON-CL4-2021-DIGITAL-EMERGING-01-30 via the project 101070144 (EuRyQa) and from the French National
Research Agency under the Investments of the Future
Program projects ANR-21-ESRE-0032 (aQCess), ANR-22-CE47-0013-02 (CLIMAQS) and ANR-23-CE30-0022-02 (SIX). NP acknowledges support from
the National Science Foundation under Grant No. DMR-2335904.
\end{acknowledgments}

\footnotetext[1]{See supplemental material including the dispersion relation in the superfluid phase with finite disorder. The supplemental material includes Refs.~\cite{dupuis2024,prokofev_ac_2013,prokofev_ac_2017}.}

\bibliography{references}

\begin{thebibliography}{34}%
\makeatletter
\providecommand \@ifxundefined [1]{%
 \@ifx{#1\undefined}
}%
\providecommand \@ifnum [1]{%
 \ifnum #1\expandafter \@firstoftwo
 \else \expandafter \@secondoftwo
 \fi
}%
\providecommand \@ifx [1]{%
 \ifx #1\expandafter \@firstoftwo
 \else \expandafter \@secondoftwo
 \fi
}%
\providecommand \natexlab [1]{#1}%
\providecommand \enquote  [1]{``#1''}%
\providecommand \bibnamefont  [1]{#1}%
\providecommand \bibfnamefont [1]{#1}%
\providecommand \citenamefont [1]{#1}%
\providecommand \href@noop [0]{\@secondoftwo}%
\providecommand \href [0]{\begingroup \@sanitize@url \@href}%
\providecommand \@href[1]{\@@startlink{#1}\@@href}%
\providecommand \@@href[1]{\endgroup#1\@@endlink}%
\providecommand \@sanitize@url [0]{\catcode `\\12\catcode `\$12\catcode `\&12\catcode `\#12\catcode `\^12\catcode `\_12\catcode `\%12\relax}%
\providecommand \@@startlink[1]{}%
\providecommand \@@endlink[0]{}%
\providecommand \url  [0]{\begingroup\@sanitize@url \@url }%
\providecommand \@url [1]{\endgroup\@href {#1}{\urlprefix }}%
\providecommand \urlprefix  [0]{URL }%
\providecommand \Eprint [0]{\href }%
\providecommand \doibase [0]{https://doi.org/}%
\providecommand \selectlanguage [0]{\@gobble}%
\providecommand \bibinfo  [0]{\@secondoftwo}%
\providecommand \bibfield  [0]{\@secondoftwo}%
\providecommand \translation [1]{[#1]}%
\providecommand \BibitemOpen [0]{}%
\providecommand \bibitemStop [0]{}%
\providecommand \bibitemNoStop [0]{.\EOS\space}%
\providecommand \EOS [0]{\spacefactor3000\relax}%
\providecommand \BibitemShut  [1]{\csname bibitem#1\endcsname}%
\let\auto@bib@innerbib\@empty
\bibitem [{\citenamefont {Giamarchi}\ and\ \citenamefont {Schulz}(1988)}]{Giamarchi1988}%
  \BibitemOpen
  \bibfield  {author} {\bibinfo {author} {\bibfnamefont {T.}~\bibnamefont {Giamarchi}}\ and\ \bibinfo {author} {\bibfnamefont {H.~J.}\ \bibnamefont {Schulz}},\ }\bibfield  {title} {\bibinfo {title} {Anderson localization and interactions in one-dimensional metals},\ }\href {https://doi.org/10.1103/PhysRevB.37.325} {\bibfield  {journal} {\bibinfo  {journal} {Phys. Rev. B}\ }\textbf {\bibinfo {volume} {37}},\ \bibinfo {pages} {325} (\bibinfo {year} {1988})}\BibitemShut {NoStop}%
\bibitem [{\citenamefont {Ristivojevic}\ \emph {et~al.}(2014)\citenamefont {Ristivojevic}, \citenamefont {Petkovi\ifmmode~\acute{c}\else \'{c}\fi{}}, \citenamefont {Le~Doussal},\ and\ \citenamefont {Giamarchi}}]{Ristivojevic2014}%
  \BibitemOpen
  \bibfield  {author} {\bibinfo {author} {\bibfnamefont {Z.}~\bibnamefont {Ristivojevic}}, \bibinfo {author} {\bibfnamefont {A.}~\bibnamefont {Petkovi\ifmmode~\acute{c}\else \'{c}\fi{}}}, \bibinfo {author} {\bibfnamefont {P.}~\bibnamefont {Le~Doussal}},\ and\ \bibinfo {author} {\bibfnamefont {T.}~\bibnamefont {Giamarchi}},\ }\bibfield  {title} {\bibinfo {title} {Superfluid/bose-glass transition in one dimension},\ }\href {https://doi.org/10.1103/PhysRevB.90.125144} {\bibfield  {journal} {\bibinfo  {journal} {Phys. Rev. B}\ }\textbf {\bibinfo {volume} {90}},\ \bibinfo {pages} {125144} (\bibinfo {year} {2014})}\BibitemShut {NoStop}%
\bibitem [{\citenamefont {Altman}\ \emph {et~al.}(2008)\citenamefont {Altman}, \citenamefont {Kafri}, \citenamefont {Polkovnikov},\ and\ \citenamefont {Refael}}]{Altman2008}%
  \BibitemOpen
  \bibfield  {author} {\bibinfo {author} {\bibfnamefont {E.}~\bibnamefont {Altman}}, \bibinfo {author} {\bibfnamefont {Y.}~\bibnamefont {Kafri}}, \bibinfo {author} {\bibfnamefont {A.}~\bibnamefont {Polkovnikov}},\ and\ \bibinfo {author} {\bibfnamefont {G.}~\bibnamefont {Refael}},\ }\bibfield  {title} {\bibinfo {title} {Insulating phases and superfluid-insulator transition of disordered boson chains},\ }\href {https://doi.org/10.1103/PhysRevLett.100.170402} {\bibfield  {journal} {\bibinfo  {journal} {Phys. Rev. Lett.}\ }\textbf {\bibinfo {volume} {100}},\ \bibinfo {pages} {170402} (\bibinfo {year} {2008})}\BibitemShut {NoStop}%
\bibitem [{\citenamefont {Altman}\ \emph {et~al.}(2010)\citenamefont {Altman}, \citenamefont {Kafri}, \citenamefont {Polkovnikov},\ and\ \citenamefont {Refael}}]{Altman2010}%
  \BibitemOpen
  \bibfield  {author} {\bibinfo {author} {\bibfnamefont {E.}~\bibnamefont {Altman}}, \bibinfo {author} {\bibfnamefont {Y.}~\bibnamefont {Kafri}}, \bibinfo {author} {\bibfnamefont {A.}~\bibnamefont {Polkovnikov}},\ and\ \bibinfo {author} {\bibfnamefont {G.}~\bibnamefont {Refael}},\ }\bibfield  {title} {\bibinfo {title} {Superfluid-insulator transition of disordered bosons in one dimension},\ }\href {https://doi.org/10.1103/PhysRevB.81.174528} {\bibfield  {journal} {\bibinfo  {journal} {Phys. Rev. B}\ }\textbf {\bibinfo {volume} {81}},\ \bibinfo {pages} {174528} (\bibinfo {year} {2010})}\BibitemShut {NoStop}%
\bibitem [{\citenamefont {Pollet}\ \emph {et~al.}(2014)\citenamefont {Pollet}, \citenamefont {Prokof'ev},\ and\ \citenamefont {Svistunov}}]{PolletNikolay2014}%
  \BibitemOpen
  \bibfield  {author} {\bibinfo {author} {\bibfnamefont {L.}~\bibnamefont {Pollet}}, \bibinfo {author} {\bibfnamefont {N.~V.}\ \bibnamefont {Prokof'ev}},\ and\ \bibinfo {author} {\bibfnamefont {B.~V.}\ \bibnamefont {Svistunov}},\ }\bibfield  {title} {\bibinfo {title} {Asymptotically exact scenario of strong-disorder criticality in one-dimensional superfluids},\ }\href {https://doi.org/10.1103/PhysRevB.89.054204} {\bibfield  {journal} {\bibinfo  {journal} {Phys. Rev. B}\ }\textbf {\bibinfo {volume} {89}},\ \bibinfo {pages} {054204} (\bibinfo {year} {2014})}\BibitemShut {NoStop}%
\bibitem [{\citenamefont {Prokof’ev}\ \emph {et~al.}(1998)\citenamefont {Prokof’ev}, \citenamefont {Svistunov},\ and\ \citenamefont {Tupitsyn}}]{prokof1998exact}%
  \BibitemOpen
  \bibfield  {author} {\bibinfo {author} {\bibfnamefont {N.~V.}\ \bibnamefont {Prokof’ev}}, \bibinfo {author} {\bibfnamefont {B.}~\bibnamefont {Svistunov}},\ and\ \bibinfo {author} {\bibfnamefont {I.}~\bibnamefont {Tupitsyn}},\ }\bibfield  {title} {\bibinfo {title} {Exact, complete, and universal continuous-time worldline monte carlo approach to the statistics of discrete quantum systems},\ }\href {https://doi.org/10.1134/1.558661} {\bibfield  {journal} {\bibinfo  {journal} {Journal of Experimental and Theoretical Physics}\ }\textbf {\bibinfo {volume} {87}},\ \bibinfo {pages} {310} (\bibinfo {year} {1998})}\BibitemShut {NoStop}%
\bibitem [{\citenamefont {Yan}\ \emph {et~al.}(2013)\citenamefont {Yan}, \citenamefont {Moses}, \citenamefont {Gadway}, \citenamefont {Covey}, \citenamefont {Hazzard}, \citenamefont {Rey}, \citenamefont {Jin},\ and\ \citenamefont {Ye}}]{Yan2013}%
  \BibitemOpen
  \bibfield  {author} {\bibinfo {author} {\bibfnamefont {B.}~\bibnamefont {Yan}}, \bibinfo {author} {\bibfnamefont {S.~A.}\ \bibnamefont {Moses}}, \bibinfo {author} {\bibfnamefont {B.}~\bibnamefont {Gadway}}, \bibinfo {author} {\bibfnamefont {J.~P.}\ \bibnamefont {Covey}}, \bibinfo {author} {\bibfnamefont {K.~R.}\ \bibnamefont {Hazzard}}, \bibinfo {author} {\bibfnamefont {A.~M.}\ \bibnamefont {Rey}}, \bibinfo {author} {\bibfnamefont {D.~S.}\ \bibnamefont {Jin}},\ and\ \bibinfo {author} {\bibfnamefont {J.}~\bibnamefont {Ye}},\ }\bibfield  {title} {\bibinfo {title} {Observation of dipolar spin-exchange interactions with lattice-confined polar molecules},\ }\href {https://doi.org/10.1038/nature12483} {\bibfield  {journal} {\bibinfo  {journal} {Nature}\ }\textbf {\bibinfo {volume} {501}},\ \bibinfo {pages} {521} (\bibinfo {year} {2013})}\BibitemShut {NoStop}%
\bibitem [{\citenamefont {Richerme}\ \emph {et~al.}(2014)\citenamefont {Richerme}, \citenamefont {Gong}, \citenamefont {Lee}, \citenamefont {Senko}, \citenamefont {Smith}, \citenamefont {Foss-Feig}, \citenamefont {Michalakis}, \citenamefont {Gorshkov},\ and\ \citenamefont {Monroe}}]{Richerme2014}%
  \BibitemOpen
  \bibfield  {author} {\bibinfo {author} {\bibfnamefont {P.}~\bibnamefont {Richerme}}, \bibinfo {author} {\bibfnamefont {Z.-X.}\ \bibnamefont {Gong}}, \bibinfo {author} {\bibfnamefont {A.}~\bibnamefont {Lee}}, \bibinfo {author} {\bibfnamefont {C.}~\bibnamefont {Senko}}, \bibinfo {author} {\bibfnamefont {J.}~\bibnamefont {Smith}}, \bibinfo {author} {\bibfnamefont {M.}~\bibnamefont {Foss-Feig}}, \bibinfo {author} {\bibfnamefont {S.}~\bibnamefont {Michalakis}}, \bibinfo {author} {\bibfnamefont {A.~V.}\ \bibnamefont {Gorshkov}},\ and\ \bibinfo {author} {\bibfnamefont {C.}~\bibnamefont {Monroe}},\ }\bibfield  {title} {\bibinfo {title} {Non-local propagation of correlations in quantum systems with long-range interactions},\ }\href {https://doi.org/10.1038/nature13450} {\bibfield  {journal} {\bibinfo  {journal} {Nature}\ }\textbf {\bibinfo {volume} {511}},\ \bibinfo {pages} {198} (\bibinfo {year} {2014})}\BibitemShut {NoStop}%
\bibitem [{\citenamefont {Jurcevic}\ \emph {et~al.}(2014)\citenamefont {Jurcevic}, \citenamefont {Lanyon}, \citenamefont {Hauke}, \citenamefont {Hempel}, \citenamefont {Zoller}, \citenamefont {Blatt},\ and\ \citenamefont {Roos}}]{Jurcevic2014}%
  \BibitemOpen
  \bibfield  {author} {\bibinfo {author} {\bibfnamefont {P.}~\bibnamefont {Jurcevic}}, \bibinfo {author} {\bibfnamefont {B.~P.}\ \bibnamefont {Lanyon}}, \bibinfo {author} {\bibfnamefont {P.}~\bibnamefont {Hauke}}, \bibinfo {author} {\bibfnamefont {C.}~\bibnamefont {Hempel}}, \bibinfo {author} {\bibfnamefont {P.}~\bibnamefont {Zoller}}, \bibinfo {author} {\bibfnamefont {R.}~\bibnamefont {Blatt}},\ and\ \bibinfo {author} {\bibfnamefont {C.~F.}\ \bibnamefont {Roos}},\ }\bibfield  {title} {\bibinfo {title} {Quasiparticle engineering and entanglement propagation in a quantum many-body system},\ }\href {https://doi.org/10.1038/nature13461} {\bibfield  {journal} {\bibinfo  {journal} {Nature}\ }\textbf {\bibinfo {volume} {511}},\ \bibinfo {pages} {202} (\bibinfo {year} {2014})}\BibitemShut {NoStop}%
\bibitem [{\citenamefont {Feng}\ \emph {et~al.}(2023)\citenamefont {Feng}, \citenamefont {Katz}, \citenamefont {Haack}, \citenamefont {Maghrebi}, \citenamefont {Gorshkov}, \citenamefont {Gong}, \citenamefont {Cetina},\ and\ \citenamefont {Monroe}}]{feng2022continuous}%
  \BibitemOpen
  \bibfield  {author} {\bibinfo {author} {\bibfnamefont {L.}~\bibnamefont {Feng}}, \bibinfo {author} {\bibfnamefont {O.}~\bibnamefont {Katz}}, \bibinfo {author} {\bibfnamefont {C.}~\bibnamefont {Haack}}, \bibinfo {author} {\bibfnamefont {M.}~\bibnamefont {Maghrebi}}, \bibinfo {author} {\bibfnamefont {A.~V.}\ \bibnamefont {Gorshkov}}, \bibinfo {author} {\bibfnamefont {Z.}~\bibnamefont {Gong}}, \bibinfo {author} {\bibfnamefont {M.}~\bibnamefont {Cetina}},\ and\ \bibinfo {author} {\bibfnamefont {C.}~\bibnamefont {Monroe}},\ }\bibfield  {title} {\bibinfo {title} {Continuous symmetry breaking in a trapped-ion spin chain},\ }\href {https://doi.org/10.1038/s41586-023-06656-7} {\bibfield  {journal} {\bibinfo  {journal} {Nature}\ }\textbf {\bibinfo {volume} {623}},\ \bibinfo {pages} {713} (\bibinfo {year} {2023})}\BibitemShut {NoStop}%
\bibitem [{\citenamefont {Zeiher}\ \emph {et~al.}(2017)\citenamefont {Zeiher}, \citenamefont {Choi}, \citenamefont {Rubio-Abadal}, \citenamefont {Pohl}, \citenamefont {van Bijnen}, \citenamefont {Bloch},\ and\ \citenamefont {Gross}}]{Zeiher2017}%
  \BibitemOpen
  \bibfield  {author} {\bibinfo {author} {\bibfnamefont {J.}~\bibnamefont {Zeiher}}, \bibinfo {author} {\bibfnamefont {J.-y.}\ \bibnamefont {Choi}}, \bibinfo {author} {\bibfnamefont {A.}~\bibnamefont {Rubio-Abadal}}, \bibinfo {author} {\bibfnamefont {T.}~\bibnamefont {Pohl}}, \bibinfo {author} {\bibfnamefont {R.}~\bibnamefont {van Bijnen}}, \bibinfo {author} {\bibfnamefont {I.}~\bibnamefont {Bloch}},\ and\ \bibinfo {author} {\bibfnamefont {C.}~\bibnamefont {Gross}},\ }\bibfield  {title} {\bibinfo {title} {Coherent many-body spin dynamics in a long-range interacting ising chain},\ }\href {https://doi.org/10.1103/PhysRevX.7.041063} {\bibfield  {journal} {\bibinfo  {journal} {Phys. Rev. X}\ }\textbf {\bibinfo {volume} {7}},\ \bibinfo {pages} {041063} (\bibinfo {year} {2017})}\BibitemShut {NoStop}%
\bibitem [{\citenamefont {Barredo}\ \emph {et~al.}(2015)\citenamefont {Barredo}, \citenamefont {Labuhn}, \citenamefont {Ravets}, \citenamefont {Lahaye}, \citenamefont {Browaeys},\ and\ \citenamefont {Adams}}]{Barredo2015}%
  \BibitemOpen
  \bibfield  {author} {\bibinfo {author} {\bibfnamefont {D.}~\bibnamefont {Barredo}}, \bibinfo {author} {\bibfnamefont {H.}~\bibnamefont {Labuhn}}, \bibinfo {author} {\bibfnamefont {S.}~\bibnamefont {Ravets}}, \bibinfo {author} {\bibfnamefont {T.}~\bibnamefont {Lahaye}}, \bibinfo {author} {\bibfnamefont {A.}~\bibnamefont {Browaeys}},\ and\ \bibinfo {author} {\bibfnamefont {C.~S.}\ \bibnamefont {Adams}},\ }\bibfield  {title} {\bibinfo {title} {Coherent excitation transfer in a spin chain of three {Rydberg} atoms},\ }\href {https://doi.org/10.1103/PhysRevLett.114.113002} {\bibfield  {journal} {\bibinfo  {journal} {Phys. Rev. Lett.}\ }\textbf {\bibinfo {volume} {114}},\ \bibinfo {pages} {113002} (\bibinfo {year} {2015})}\BibitemShut {NoStop}%
\bibitem [{\citenamefont {Orioli}\ \emph {et~al.}(2018)\citenamefont {Orioli}, \citenamefont {Signoles}, \citenamefont {Wildhagen}, \citenamefont {G{\"u}nter}, \citenamefont {Berges}, \citenamefont {Whitlock},\ and\ \citenamefont {Weidem{\"u}ller}}]{Orioli2018}%
  \BibitemOpen
  \bibfield  {author} {\bibinfo {author} {\bibfnamefont {A.~P.}\ \bibnamefont {Orioli}}, \bibinfo {author} {\bibfnamefont {A.}~\bibnamefont {Signoles}}, \bibinfo {author} {\bibfnamefont {H.}~\bibnamefont {Wildhagen}}, \bibinfo {author} {\bibfnamefont {G.}~\bibnamefont {G{\"u}nter}}, \bibinfo {author} {\bibfnamefont {J.}~\bibnamefont {Berges}}, \bibinfo {author} {\bibfnamefont {S.}~\bibnamefont {Whitlock}},\ and\ \bibinfo {author} {\bibfnamefont {M.}~\bibnamefont {Weidem{\"u}ller}},\ }\bibfield  {title} {\bibinfo {title} {Relaxation of an {{Isolated Dipolar-Interacting Rydberg Quantum Spin System}}},\ }\href {https://doi.org/10.1103/PhysRevLett.120.063601} {\bibfield  {journal} {\bibinfo  {journal} {Physical Review Letters}\ }\textbf {\bibinfo {volume} {120}},\ \bibinfo {pages} {063601} (\bibinfo {year} {2018})}\BibitemShut {NoStop}%
\bibitem [{\citenamefont {Browaeys}\ and\ \citenamefont {Lahaye}(2020)}]{browaeys_lahaye_2020}%
  \BibitemOpen
  \bibfield  {author} {\bibinfo {author} {\bibfnamefont {A.}~\bibnamefont {Browaeys}}\ and\ \bibinfo {author} {\bibfnamefont {T.}~\bibnamefont {Lahaye}},\ }\bibfield  {title} {\bibinfo {title} {Many-body physics with individually controlled {Rydberg} atoms},\ }\href {https://doi.org/10.1038/s41567-019-0733-z} {\bibfield  {journal} {\bibinfo  {journal} {Nature Physics}\ }\textbf {\bibinfo {volume} {16}},\ \bibinfo {pages} {132–142} (\bibinfo {year} {2020})}\BibitemShut {NoStop}%
\bibitem [{\citenamefont {Semeghini}\ \emph {et~al.}(2021)\citenamefont {Semeghini}, \citenamefont {Levine}, \citenamefont {Keesling}, \citenamefont {Ebadi}, \citenamefont {Wang}, \citenamefont {Bluvstein}, \citenamefont {Verresen}, \citenamefont {Pichler}, \citenamefont {Kalinowski}, \citenamefont {Samajdar}, \citenamefont {Omran}, \citenamefont {Sachdev}, \citenamefont {Vishwanath}, \citenamefont {Greiner}, \citenamefont {Vuletić},\ and\ \citenamefont {Lukin}}]{Semeghini2021}%
  \BibitemOpen
  \bibfield  {author} {\bibinfo {author} {\bibfnamefont {G.}~\bibnamefont {Semeghini}}, \bibinfo {author} {\bibfnamefont {H.}~\bibnamefont {Levine}}, \bibinfo {author} {\bibfnamefont {A.}~\bibnamefont {Keesling}}, \bibinfo {author} {\bibfnamefont {S.}~\bibnamefont {Ebadi}}, \bibinfo {author} {\bibfnamefont {T.~T.}\ \bibnamefont {Wang}}, \bibinfo {author} {\bibfnamefont {D.}~\bibnamefont {Bluvstein}}, \bibinfo {author} {\bibfnamefont {R.}~\bibnamefont {Verresen}}, \bibinfo {author} {\bibfnamefont {H.}~\bibnamefont {Pichler}}, \bibinfo {author} {\bibfnamefont {M.}~\bibnamefont {Kalinowski}}, \bibinfo {author} {\bibfnamefont {R.}~\bibnamefont {Samajdar}}, \bibinfo {author} {\bibfnamefont {A.}~\bibnamefont {Omran}}, \bibinfo {author} {\bibfnamefont {S.}~\bibnamefont {Sachdev}}, \bibinfo {author} {\bibfnamefont {A.}~\bibnamefont {Vishwanath}}, \bibinfo {author} {\bibfnamefont {M.}~\bibnamefont {Greiner}}, \bibinfo {author} {\bibfnamefont {V.}~\bibnamefont {Vuletić}},\ and\ \bibinfo {author} {\bibfnamefont
  {M.~D.}\ \bibnamefont {Lukin}},\ }\bibfield  {title} {\bibinfo {title} {Probing topological spin liquids on a programmable quantum simulator},\ }\href {https://doi.org/10.1126/science.abi8794} {\bibfield  {journal} {\bibinfo  {journal} {Science}\ }\textbf {\bibinfo {volume} {374}},\ \bibinfo {pages} {1242} (\bibinfo {year} {2021})}\BibitemShut {NoStop}%
\bibitem [{\citenamefont {Chen}\ \emph {et~al.}(2023)\citenamefont {Chen}, \citenamefont {Bornet}, \citenamefont {Bintz}, \citenamefont {Emperauger}, \citenamefont {Leclerc}, \citenamefont {Liu}, \citenamefont {Scholl}, \citenamefont {Barredo}, \citenamefont {Hauschild}, \citenamefont {Chatterjee}, \citenamefont {Schuler}, \citenamefont {Läuchli}, \citenamefont {Zaletel}, \citenamefont {Lahaye}, \citenamefont {Yao},\ and\ \citenamefont {Browaeys}}]{chen2023continuous}%
  \BibitemOpen
  \bibfield  {author} {\bibinfo {author} {\bibfnamefont {C.}~\bibnamefont {Chen}}, \bibinfo {author} {\bibfnamefont {G.}~\bibnamefont {Bornet}}, \bibinfo {author} {\bibfnamefont {M.}~\bibnamefont {Bintz}}, \bibinfo {author} {\bibfnamefont {G.}~\bibnamefont {Emperauger}}, \bibinfo {author} {\bibfnamefont {L.}~\bibnamefont {Leclerc}}, \bibinfo {author} {\bibfnamefont {V.~S.}\ \bibnamefont {Liu}}, \bibinfo {author} {\bibfnamefont {P.}~\bibnamefont {Scholl}}, \bibinfo {author} {\bibfnamefont {D.}~\bibnamefont {Barredo}}, \bibinfo {author} {\bibfnamefont {J.}~\bibnamefont {Hauschild}}, \bibinfo {author} {\bibfnamefont {S.}~\bibnamefont {Chatterjee}}, \bibinfo {author} {\bibfnamefont {M.}~\bibnamefont {Schuler}}, \bibinfo {author} {\bibfnamefont {A.~M.}\ \bibnamefont {Läuchli}}, \bibinfo {author} {\bibfnamefont {M.~P.}\ \bibnamefont {Zaletel}}, \bibinfo {author} {\bibfnamefont {T.}~\bibnamefont {Lahaye}}, \bibinfo {author} {\bibfnamefont {N.~Y.}\ \bibnamefont {Yao}},\ and\ \bibinfo {author} {\bibfnamefont
  {A.}~\bibnamefont {Browaeys}},\ }\bibfield  {title} {\bibinfo {title} {Continuous symmetry breaking in a two-dimensional {Rydberg} array},\ }\href {https://doi.org/10.1038/s41586-023-05859-2} {\bibfield  {journal} {\bibinfo  {journal} {Nature}\ }\textbf {\bibinfo {volume} {616}},\ \bibinfo {pages} {691} (\bibinfo {year} {2023})}\BibitemShut {NoStop}%
\bibitem [{\citenamefont {de~Léséleuc}\ \emph {et~al.}(2019)\citenamefont {de~Léséleuc}, \citenamefont {Lienhard}, \citenamefont {Scholl}, \citenamefont {Barredo}, \citenamefont {Weber}, \citenamefont {Lang}, \citenamefont {Büchler}, \citenamefont {Lahaye},\ and\ \citenamefont {Browaeys}}]{Leseleuc2019}%
  \BibitemOpen
  \bibfield  {author} {\bibinfo {author} {\bibfnamefont {S.}~\bibnamefont {de~Léséleuc}}, \bibinfo {author} {\bibfnamefont {V.}~\bibnamefont {Lienhard}}, \bibinfo {author} {\bibfnamefont {P.}~\bibnamefont {Scholl}}, \bibinfo {author} {\bibfnamefont {D.}~\bibnamefont {Barredo}}, \bibinfo {author} {\bibfnamefont {S.}~\bibnamefont {Weber}}, \bibinfo {author} {\bibfnamefont {N.}~\bibnamefont {Lang}}, \bibinfo {author} {\bibfnamefont {H.~P.}\ \bibnamefont {Büchler}}, \bibinfo {author} {\bibfnamefont {T.}~\bibnamefont {Lahaye}},\ and\ \bibinfo {author} {\bibfnamefont {A.}~\bibnamefont {Browaeys}},\ }\bibfield  {title} {\bibinfo {title} {Observation of a symmetry-protected topological phase of interacting bosons with {Rydberg} atoms},\ }\href {https://doi.org/10.1126/science.aav9105} {\bibfield  {journal} {\bibinfo  {journal} {Science}\ }\textbf {\bibinfo {volume} {365}},\ \bibinfo {pages} {775} (\bibinfo {year} {2019})}\BibitemShut {NoStop}%
\bibitem [{\citenamefont {Fr\'erot}\ \emph {et~al.}(2017)\citenamefont {Fr\'erot}, \citenamefont {Naldesi},\ and\ \citenamefont {Roscilde}}]{Roscilde2017}%
  \BibitemOpen
  \bibfield  {author} {\bibinfo {author} {\bibfnamefont {I.}~\bibnamefont {Fr\'erot}}, \bibinfo {author} {\bibfnamefont {P.}~\bibnamefont {Naldesi}},\ and\ \bibinfo {author} {\bibfnamefont {T.}~\bibnamefont {Roscilde}},\ }\bibfield  {title} {\bibinfo {title} {Entanglement and fluctuations in the {XXZ} model with power-law interactions},\ }\href {https://doi.org/10.1103/PhysRevB.95.245111} {\bibfield  {journal} {\bibinfo  {journal} {Phys. Rev. B}\ }\textbf {\bibinfo {volume} {95}},\ \bibinfo {pages} {245111} (\bibinfo {year} {2017})}\BibitemShut {NoStop}%
\bibitem [{\citenamefont {Maghrebi}\ \emph {et~al.}(2017)\citenamefont {Maghrebi}, \citenamefont {Gong},\ and\ \citenamefont {Gorshkov}}]{maghrebi_gong_gorshkov_2017}%
  \BibitemOpen
  \bibfield  {author} {\bibinfo {author} {\bibfnamefont {M.~F.}\ \bibnamefont {Maghrebi}}, \bibinfo {author} {\bibfnamefont {Z.-X.}\ \bibnamefont {Gong}},\ and\ \bibinfo {author} {\bibfnamefont {A.~V.}\ \bibnamefont {Gorshkov}},\ }\bibfield  {title} {\bibinfo {title} {Continuous symmetry breaking in 1d long-range interacting quantum systems},\ }\href {https://doi.org/10.1103/PhysRevLett.119.023001} {\bibfield  {journal} {\bibinfo  {journal} {Phys. Rev. Lett.}\ }\textbf {\bibinfo {volume} {119}},\ \bibinfo {pages} {023001} (\bibinfo {year} {2017})}\BibitemShut {NoStop}%
\bibitem [{\citenamefont {Prokof’Ev}\ and\ \citenamefont {Svistunov}(2013)}]{prokofev_ac_2013}%
  \BibitemOpen
  \bibfield  {author} {\bibinfo {author} {\bibfnamefont {N.~V.}\ \bibnamefont {Prokof’Ev}}\ and\ \bibinfo {author} {\bibfnamefont {B.~V.}\ \bibnamefont {Svistunov}},\ }\bibfield  {title} {\bibinfo {title} {Spectral analysis by the method of consistent constraints},\ }\href {https://doi.org/10.1134/s002136401311009x} {\bibfield  {journal} {\bibinfo  {journal} {JETP Letters}\ }\textbf {\bibinfo {volume} {97}},\ \bibinfo {pages} {649–653} (\bibinfo {year} {2013})}\BibitemShut {NoStop}%
\bibitem [{\citenamefont {Goulko}\ \emph {et~al.}(2017)\citenamefont {Goulko}, \citenamefont {Mishchenko}, \citenamefont {Pollet}, \citenamefont {Prokof'ev},\ and\ \citenamefont {Svistunov}}]{prokofev_ac_2017}%
  \BibitemOpen
  \bibfield  {author} {\bibinfo {author} {\bibfnamefont {O.}~\bibnamefont {Goulko}}, \bibinfo {author} {\bibfnamefont {A.~S.}\ \bibnamefont {Mishchenko}}, \bibinfo {author} {\bibfnamefont {L.}~\bibnamefont {Pollet}}, \bibinfo {author} {\bibfnamefont {N.}~\bibnamefont {Prokof'ev}},\ and\ \bibinfo {author} {\bibfnamefont {B.}~\bibnamefont {Svistunov}},\ }\bibfield  {title} {\bibinfo {title} {Numerical analytic continuation: Answers to well-posed questions},\ }\href {https://doi.org/10.1103/PhysRevB.95.014102} {\bibfield  {journal} {\bibinfo  {journal} {Phys. Rev. B}\ }\textbf {\bibinfo {volume} {95}},\ \bibinfo {pages} {014102} (\bibinfo {year} {2017})}\BibitemShut {NoStop}%
\bibitem [{Note1()}]{Note1}%
  \BibitemOpen
  \bibinfo {note} {See supplemental material including the dispersion relation in the superfluid phase with finite disorder. The supplemental material includes Refs.~\cite {dupuis2024,prokofev_ac_2013,prokofev_ac_2017}.}\BibitemShut {Stop}%
\bibitem [{\citenamefont {Giamarchi}(2003)}]{GiamarchiBook}%
  \BibitemOpen
  \bibfield  {author} {\bibinfo {author} {\bibfnamefont {T.}~\bibnamefont {Giamarchi}},\ }\href {https://doi.org/10.1093/acprof:oso/9780198525004.001.0001} {\emph {\bibinfo {title} {Quantum {Physics} in {One} {Dimension}}}}\ (\bibinfo  {publisher} {Oxford University Press},\ \bibinfo {year} {2003})\BibitemShut {NoStop}%
\bibitem [{\citenamefont {Newman}\ and\ \citenamefont {Barkema}(1999)}]{newmanb99}%
  \BibitemOpen
  \bibfield  {author} {\bibinfo {author} {\bibfnamefont {M.~E.~J.}\ \bibnamefont {Newman}}\ and\ \bibinfo {author} {\bibfnamefont {G.~T.}\ \bibnamefont {Barkema}},\ }\href {https://doi.org/10.1093/oso/9780198517962.001.0001} {\emph {\bibinfo {title} {Monte {Carlo} {Methods} in {Statistical} {Physics}}}}\ (\bibinfo  {publisher} {Oxford University Press, Oxford},\ \bibinfo {year} {1999})\BibitemShut {NoStop}%
\bibitem [{\citenamefont {Nelder}\ and\ \citenamefont {Mead}(1965)}]{Nelder1965}%
  \BibitemOpen
  \bibfield  {author} {\bibinfo {author} {\bibfnamefont {J.~A.}\ \bibnamefont {Nelder}}\ and\ \bibinfo {author} {\bibfnamefont {R.}~\bibnamefont {Mead}},\ }\bibfield  {title} {\bibinfo {title} {{A Simplex Method for Function Minimization}},\ }\href {https://doi.org/10.1093/comjnl/7.4.308} {\bibfield  {journal} {\bibinfo  {journal} {The Computer Journal}\ }\textbf {\bibinfo {volume} {7}},\ \bibinfo {pages} {308} (\bibinfo {year} {1965})}\BibitemShut {NoStop}%
\bibitem [{\citenamefont {Kawashima}\ and\ \citenamefont {Ito}(1993)}]{Kawashima1993}%
  \BibitemOpen
  \bibfield  {author} {\bibinfo {author} {\bibfnamefont {N.}~\bibnamefont {Kawashima}}\ and\ \bibinfo {author} {\bibfnamefont {N.}~\bibnamefont {Ito}},\ }\bibfield  {title} {\bibinfo {title} {Critical behavior of the three-dimensional {±J} model in a magnetic field},\ }\href {https://doi.org/10.1143/JPSJ.62.435} {\bibfield  {journal} {\bibinfo  {journal} {Journal of the Physical Society of Japan}\ }\textbf {\bibinfo {volume} {62}},\ \bibinfo {pages} {435} (\bibinfo {year} {1993})}\BibitemShut {NoStop}%
\bibitem [{\citenamefont {Houdayer}\ and\ \citenamefont {Hartmann}(2004)}]{Houdayer2004}%
  \BibitemOpen
  \bibfield  {author} {\bibinfo {author} {\bibfnamefont {J.}~\bibnamefont {Houdayer}}\ and\ \bibinfo {author} {\bibfnamefont {A.~K.}\ \bibnamefont {Hartmann}},\ }\bibfield  {title} {\bibinfo {title} {Low-temperature behavior of two-dimensional gaussian ising spin glasses},\ }\href {https://doi.org/10.1103/PhysRevB.70.014418} {\bibfield  {journal} {\bibinfo  {journal} {Phys. Rev. B}\ }\textbf {\bibinfo {volume} {70}},\ \bibinfo {pages} {014418} (\bibinfo {year} {2004})}\BibitemShut {NoStop}%
\bibitem [{\citenamefont {Harris}(1974)}]{harris1974effect}%
  \BibitemOpen
  \bibfield  {author} {\bibinfo {author} {\bibfnamefont {A.~B.}\ \bibnamefont {Harris}},\ }\bibfield  {title} {\bibinfo {title} {Effect of random defects on the critical behaviour of {Ising} models},\ }\href {https://doi.org/10.1088/0022-3719/7/9/009} {\bibfield  {journal} {\bibinfo  {journal} {Journal of Physics C: Solid State Physics}\ }\textbf {\bibinfo {volume} {7}},\ \bibinfo {pages} {1671} (\bibinfo {year} {1974})}\BibitemShut {NoStop}%
\bibitem [{\citenamefont {Chayes}\ \emph {et~al.}(1986)\citenamefont {Chayes}, \citenamefont {Chayes}, \citenamefont {Fisher},\ and\ \citenamefont {Spencer}}]{Chayes1986}%
  \BibitemOpen
  \bibfield  {author} {\bibinfo {author} {\bibfnamefont {J.~T.}\ \bibnamefont {Chayes}}, \bibinfo {author} {\bibfnamefont {L.}~\bibnamefont {Chayes}}, \bibinfo {author} {\bibfnamefont {D.~S.}\ \bibnamefont {Fisher}},\ and\ \bibinfo {author} {\bibfnamefont {T.}~\bibnamefont {Spencer}},\ }\bibfield  {title} {\bibinfo {title} {Finite-size scaling and correlation lengths for disordered systems},\ }\href {https://doi.org/10.1103/PhysRevLett.57.2999} {\bibfield  {journal} {\bibinfo  {journal} {Phys. Rev. Lett.}\ }\textbf {\bibinfo {volume} {57}},\ \bibinfo {pages} {2999} (\bibinfo {year} {1986})}\BibitemShut {NoStop}%
\bibitem [{\citenamefont {Sbierski}\ \emph {et~al.}(2023)\citenamefont {Sbierski}, \citenamefont {Bintz}, \citenamefont {Chatterjee}, \citenamefont {Schuler}, \citenamefont {Yao},\ and\ \citenamefont {Pollet}}]{sbierski2023magnetism}%
  \BibitemOpen
  \bibfield  {author} {\bibinfo {author} {\bibfnamefont {B.}~\bibnamefont {Sbierski}}, \bibinfo {author} {\bibfnamefont {M.}~\bibnamefont {Bintz}}, \bibinfo {author} {\bibfnamefont {S.}~\bibnamefont {Chatterjee}}, \bibinfo {author} {\bibfnamefont {M.}~\bibnamefont {Schuler}}, \bibinfo {author} {\bibfnamefont {N.~Y.}\ \bibnamefont {Yao}},\ and\ \bibinfo {author} {\bibfnamefont {L.}~\bibnamefont {Pollet}},\ }\href@noop {} {\bibinfo {title} {Magnetism in the two-dimensional dipolar {XY} model}} (\bibinfo {year} {2023}),\ \Eprint {https://arxiv.org/abs/2305.03673} {arXiv:2305.03673} \BibitemShut {NoStop}%
\bibitem [{\citenamefont {Diessel}\ \emph {et~al.}(2023)\citenamefont {Diessel}, \citenamefont {Diehl}, \citenamefont {Defenu}, \citenamefont {Rosch},\ and\ \citenamefont {Chiocchetta}}]{diessel_2023}%
  \BibitemOpen
  \bibfield  {author} {\bibinfo {author} {\bibfnamefont {O.~K.}\ \bibnamefont {Diessel}}, \bibinfo {author} {\bibfnamefont {S.}~\bibnamefont {Diehl}}, \bibinfo {author} {\bibfnamefont {N.}~\bibnamefont {Defenu}}, \bibinfo {author} {\bibfnamefont {A.}~\bibnamefont {Rosch}},\ and\ \bibinfo {author} {\bibfnamefont {A.}~\bibnamefont {Chiocchetta}},\ }\bibfield  {title} {\bibinfo {title} {Generalized higgs mechanism in long-range-interacting quantum systems},\ }\href {https://doi.org/10.1103/PhysRevResearch.5.033038} {\bibfield  {journal} {\bibinfo  {journal} {Phys. Rev. Res.}\ }\textbf {\bibinfo {volume} {5}},\ \bibinfo {pages} {033038} (\bibinfo {year} {2023})}\BibitemShut {NoStop}%
\bibitem [{\citenamefont {Defenu}\ \emph {et~al.}(2023)\citenamefont {Defenu}, \citenamefont {Donner}, \citenamefont {Macr\`{\i}}, \citenamefont {Pagano}, \citenamefont {Ruffo},\ and\ \citenamefont {Trombettoni}}]{DefenuLongRange2023}%
  \BibitemOpen
  \bibfield  {author} {\bibinfo {author} {\bibfnamefont {N.}~\bibnamefont {Defenu}}, \bibinfo {author} {\bibfnamefont {T.}~\bibnamefont {Donner}}, \bibinfo {author} {\bibfnamefont {T.}~\bibnamefont {Macr\`{\i}}}, \bibinfo {author} {\bibfnamefont {G.}~\bibnamefont {Pagano}}, \bibinfo {author} {\bibfnamefont {S.}~\bibnamefont {Ruffo}},\ and\ \bibinfo {author} {\bibfnamefont {A.}~\bibnamefont {Trombettoni}},\ }\bibfield  {title} {\bibinfo {title} {Long-range interacting quantum systems},\ }\href {https://doi.org/10.1103/RevModPhys.95.035002} {\bibfield  {journal} {\bibinfo  {journal} {Rev. Mod. Phys.}\ }\textbf {\bibinfo {volume} {95}},\ \bibinfo {pages} {035002} (\bibinfo {year} {2023})}\BibitemShut {NoStop}%
\bibitem [{\citenamefont {Masella}\ \emph {et~al.}(2022)\citenamefont {Masella}, \citenamefont {Prokof'ev},\ and\ \citenamefont {Pupillo}}]{Masella2022}%
  \BibitemOpen
  \bibfield  {author} {\bibinfo {author} {\bibfnamefont {G.}~\bibnamefont {Masella}}, \bibinfo {author} {\bibfnamefont {N.~V.}\ \bibnamefont {Prokof'ev}},\ and\ \bibinfo {author} {\bibfnamefont {G.}~\bibnamefont {Pupillo}},\ }\bibfield  {title} {\bibinfo {title} {Anti-drude metal of bosons},\ }\href {https://doi.org/10.1038/s41467-022-29708-4} {\bibfield  {journal} {\bibinfo  {journal} {Nature Communications}\ }\textbf {\bibinfo {volume} {13}},\ \bibinfo {pages} {2113} (\bibinfo {year} {2022})}\BibitemShut {NoStop}%
\bibitem [{\citenamefont {Dupuis}(2024)}]{dupuis2024}%
  \BibitemOpen
  \bibfield  {author} {\bibinfo {author} {\bibfnamefont {N.}~\bibnamefont {Dupuis}},\ }\bibfield  {title} {\bibinfo {title} {Superfluid--bose-glass transition in a system of disordered bosons with long-range hopping in one dimension},\ }\href {https://doi.org/10.1103/PhysRevA.110.033315} {\bibfield  {journal} {\bibinfo  {journal} {Phys. Rev. A}\ }\textbf {\bibinfo {volume} {110}},\ \bibinfo {pages} {033315} (\bibinfo {year} {2024})}\BibitemShut {NoStop}%
\end{thebibliography}%


\begin{thebibliography}{3}%
\makeatletter
\providecommand \@ifxundefined [1]{%
 \@ifx{#1\undefined}
}%
\providecommand \@ifnum [1]{%
 \ifnum #1\expandafter \@firstoftwo
 \else \expandafter \@secondoftwo
 \fi
}%
\providecommand \@ifx [1]{%
 \ifx #1\expandafter \@firstoftwo
 \else \expandafter \@secondoftwo
 \fi
}%
\providecommand \natexlab [1]{#1}%
\providecommand \enquote  [1]{``#1''}%
\providecommand \bibnamefont  [1]{#1}%
\providecommand \bibfnamefont [1]{#1}%
\providecommand \citenamefont [1]{#1}%
\providecommand \href@noop [0]{\@secondoftwo}%
\providecommand \href [0]{\begingroup \@sanitize@url \@href}%
\providecommand \@href[1]{\@@startlink{#1}\@@href}%
\providecommand \@@href[1]{\endgroup#1\@@endlink}%
\providecommand \@sanitize@url [0]{\catcode `\\12\catcode `\$12\catcode `\&12\catcode `\#12\catcode `\^12\catcode `\_12\catcode `\%12\relax}%
\providecommand \@@startlink[1]{}%
\providecommand \@@endlink[0]{}%
\providecommand \url  [0]{\begingroup\@sanitize@url \@url }%
\providecommand \@url [1]{\endgroup\@href {#1}{\urlprefix }}%
\providecommand \urlprefix  [0]{URL }%
\providecommand \Eprint [0]{\href }%
\providecommand \doibase [0]{https://doi.org/}%
\providecommand \selectlanguage [0]{\@gobble}%
\providecommand \bibinfo  [0]{\@secondoftwo}%
\providecommand \bibfield  [0]{\@secondoftwo}%
\providecommand \translation [1]{[#1]}%
\providecommand \BibitemOpen [0]{}%
\providecommand \bibitemStop [0]{}%
\providecommand \bibitemNoStop [0]{.\EOS\space}%
\providecommand \EOS [0]{\spacefactor3000\relax}%
\providecommand \BibitemShut  [1]{\csname bibitem#1\endcsname}%
\let\auto@bib@innerbib\@empty
\bibitem [{\citenamefont {Dupuis}(2024)}]{dupuis2024}%
  \BibitemOpen
  \bibfield  {author} {\bibinfo {author} {\bibfnamefont {N.}~\bibnamefont {Dupuis}},\ }\href {https://doi.org/10.1103/PhysRevA.110.033315} {\bibfield  {journal} {\bibinfo  {journal} {Phys. Rev. A}\ }\textbf {\bibinfo {volume} {110}},\ \bibinfo {pages} {033315} (\bibinfo {year} {2024})}\BibitemShut {NoStop}%
\bibitem [{\citenamefont {Prokof’Ev}\ and\ \citenamefont {Svistunov}(2013)}]{prokofev_ac_2013}%
  \BibitemOpen
  \bibfield  {author} {\bibinfo {author} {\bibfnamefont {N.~V.}\ \bibnamefont {Prokof’Ev}}\ and\ \bibinfo {author} {\bibfnamefont {B.~V.}\ \bibnamefont {Svistunov}},\ }\href {https://doi.org/10.1134/s002136401311009x} {\bibfield  {journal} {\bibinfo  {journal} {JETP Letters}\ }\textbf {\bibinfo {volume} {97}},\ \bibinfo {pages} {649–653} (\bibinfo {year} {2013})}\BibitemShut {NoStop}%
\bibitem [{\citenamefont {Goulko}\ \emph {et~al.}(2017)\citenamefont {Goulko}, \citenamefont {Mishchenko}, \citenamefont {Pollet}, \citenamefont {Prokof'ev},\ and\ \citenamefont {Svistunov}}]{prokofev_ac_2017}%
  \BibitemOpen
  \bibfield  {author} {\bibinfo {author} {\bibfnamefont {O.}~\bibnamefont {Goulko}}, \bibinfo {author} {\bibfnamefont {A.~S.}\ \bibnamefont {Mishchenko}}, \bibinfo {author} {\bibfnamefont {L.}~\bibnamefont {Pollet}}, \bibinfo {author} {\bibfnamefont {N.}~\bibnamefont {Prokof'ev}},\ and\ \bibinfo {author} {\bibfnamefont {B.}~\bibnamefont {Svistunov}},\ }\href {https://doi.org/10.1103/PhysRevB.95.014102} {\bibfield  {journal} {\bibinfo  {journal} {Phys. Rev. B}\ }\textbf {\bibinfo {volume} {95}},\ \bibinfo {pages} {014102} (\bibinfo {year} {2017})}\BibitemShut {NoStop}%
\end{thebibliography}%

\end{document}


\title{Supplemental Material for ``Scale-invariant phase transition of disordered bosons in one dimension''}

\author{Tanul Gupta}
\affiliation{University of Strasbourg and CNRS, CESQ and ISIS (UMR 7006), aQCess, 67000 Strasbourg, France}

\author{Guido~Masella}
\affiliation{QPerfect, 23 Rue du Loess, 67000 Strasbourg, France}

\author{Francesco~Mattiotti}
\affiliation{University of Strasbourg and CNRS, CESQ and ISIS (UMR 7006), aQCess, 67000 Strasbourg, France}

\author{Nikolay V. Prokof'ev}
\affiliation{Department of Physics, University of Massachusetts, Amherst, MA 01003, USA}

\author{Guido~Pupillo}
\affiliation{University of Strasbourg and CNRS, CESQ and ISIS (UMR 7006), aQCess, 67000 Strasbourg, France}
\affiliation{QPerfect, 23 Rue du Loess, 67000 Strasbourg, France}
\maketitle

In this work, we study a one-dimensional disordered lattice boson model with hopping amplitude decaying with distance as $1/r^\alpha$ using large scale quantum Monte-Carlo simulations. 
A recent work \cite{dupuis2024} based on approximate functional renormalization group methods (submitted to the archive after our work was published online) proposes that in the stable superfluid phase at finite disorder strength ($W/t > 0$), a density mode with linear dispersion (dynamical exponent $z = 1$) should emerge and the superfluid–Bose-glass transition follows the BKT universality class for any $\alpha$, in contradiction to our results. In this Supplemental Material we check the predictions of \cite{dupuis2024}. Our analysis based on large scale numerical simulations confirms the picture that we present in the main text.


\section{Dispersion relation in the superfluid phase with finite disorder}
The study presented in \cite{dupuis2024} proposes, using approximate functional renormalization group methods, that in the stable superfluid phase a density mode with linear dispersion (dynamical exponent $z = 1$) should emerge, and that the superfluid–Bose-glass transition follows the BKT universality class. To test these predictions, we performed large scale quantum Monte Carlo simulations for $\alpha=2.5$ and disorder strength $W/t=2.0$ to determine the dispersion relation by numerically analyzing the spectral peaks after performing an analytic continuation of the imaginary-frequency dynamic structure factor \cite{prokofev_ac_2013, prokofev_ac_2017}.

\begin{figure*}[h]
\centering
\includegraphics[width=0.45\linewidth]{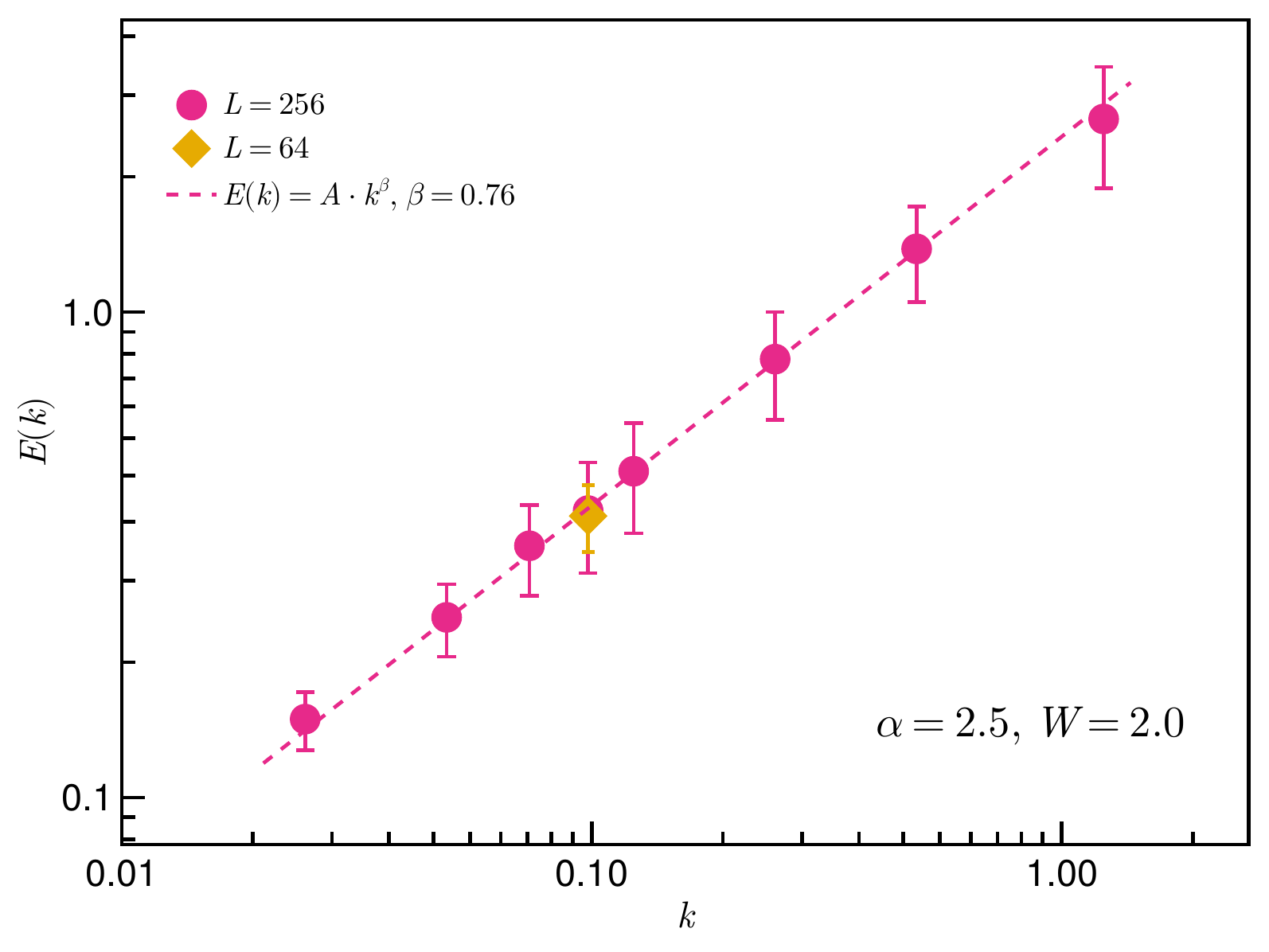}
    \caption{Dispersion relation $E(k)$ vs $k$ for $\alpha=2.5$ in the superfluid phase at finite disorder strength $W/t=2.0$.
    }
    \label{fig:EkDisorder}
\end{figure*}

Our results, presented in Figure \ref{fig:EkDisorder} of this Supplemental Material, demonstrate that the spectrum is non-linear in $k$ for $\alpha =2.5$ at $W=2.0$ and the data is well described by $E(k)\sim k^{z_*}$, with $z_* \simeq 0.76$. We checked for finite size effects (e.g., $L=64$ in the figure), founding no significant impact. These results contradict the theoretical predictions of linear dispersion proposed in \cite{dupuis2024}.

\bibliography{references}
\bibliographystyle{apsrev4-2}